\documentclass[aps,prl,twocolumn,superscriptaddress]{revtex4-1}

\usepackage{hyperref}
\usepackage{graphicx}

\newcommand*{\TN}{\ensuremath{T_{\text N}}}
\newcommand*{\wn}{\ensuremath{~\text{cm}^{-1}}}

\hypersetup{pdfborder= 0 0 0,colorlinks=false,pdftitle={Titel},pdfauthor={Christian Kant},pdfcreator={pdfLatex} }

\begin{document}

\title{Universal exchange-driven phonon splitting in antiferromagnets}

\date{\today}

\author{Ch.~Kant}
\author{M.~Schmidt}
\author{Zhe Wang}
\author{F.~Mayr}
\affiliation{Experimental Physics~V, Center for Electronic
Correlations and Magnetism, University of Augsburg, 86135~Augsburg,
Germany}
\author{V.~Tsurkan}
\affiliation{Experimental Physics~V, Center for Electronic
Correlations and Magnetism, University of Augsburg, 86135~Augsburg,
Germany} \affiliation{Institute of Applied Physics, Academy of
Sciences of Moldova, MD-2028~Chi\c{s}in\u{a}u, Republic of Moldova}
\author{J.~Deisenhofer}
\email{Corresponding author:
joachim.deisenhofer@physik.uni-augsburg.de}
\author{A.~Loidl}
\affiliation{Experimental Physics~V, Center for Electronic
Correlations and Magnetism, University of Augsburg, 86135~Augsburg,
Germany}

\begin{abstract}
We report a linear dependence of the phonon splitting $\Delta\omega$
on the non-dominant exchange coupling constant $J_{nd}$ in the
antiferromagnetic transition-metal monoxides MnO, FeO, CoO, NiO, and
in the frustrated antiferromagnetic oxide spinels CdCr$_{2}$O$_{4}$,
MgCr$_{2}$O$_{4}$, and ZnCr$_{2}$O$_{4}$. It directly confirms the
theoretical prediction of an exchange induced splitting of the
zone-centre optical phonon for the monoxides and explains the
magnitude and the change of sign of the phonon splitting on changing
the sign of the non-dominant exchange also in the frustrated oxide
spinels. The experimentally found linear relation $\hbar\Delta\omega
= \beta J_{nd} S^2$ with slope $\beta$ = 3.7 describes the splitting
for both systems and agrees with the observations in the
antiferromagnets KCoF$_3$ and KNiF$_3$ with perovskite structure and
negligible next-nearest neighbour coupling. The common behavior
found for very different classes of cubic antiferromagnets suggests
a universal dependence of the exchange-induced phonon splitting at
the antiferromagnetic transition on the non-dominant exchange
coupling.
\end{abstract}

%\pacs{}

\maketitle

The interplay of magnetism and the underlying crystal lattice is a
topical issue of condensed-matter physics. This spin-phonon coupling
can relieve frustration via a spin-driven Jahn-Teller effect in
frustrated magnets \cite{yamashit00,tchernys02}, lead to novel
excitations such as electromagnons in multiferroics
\cite{pimenov06,katsura07}, and can even bear the potential for
future applications via magneto-dielectric effects \cite{lawes05}.
For transition-metal monoxides (TMMOs) a magnetism-induced
anisotropy in the lattice response was predicted theoretically
\cite{massidda99}. This approach has been extended to other material
classes such as Cr based spinels, which are hallmark systems for
highly frustrated magnets \cite{lee02,chung05,tomiyasu04,tsurkan11},
where spin-phonon coupling leads to a splitting of characteristic
phonon modes \cite{sushkov05,fennie06,aguilar08,rudolf09a,kant09a}.

\begin{figure}[t]
\includegraphics[width=0.45\textwidth]{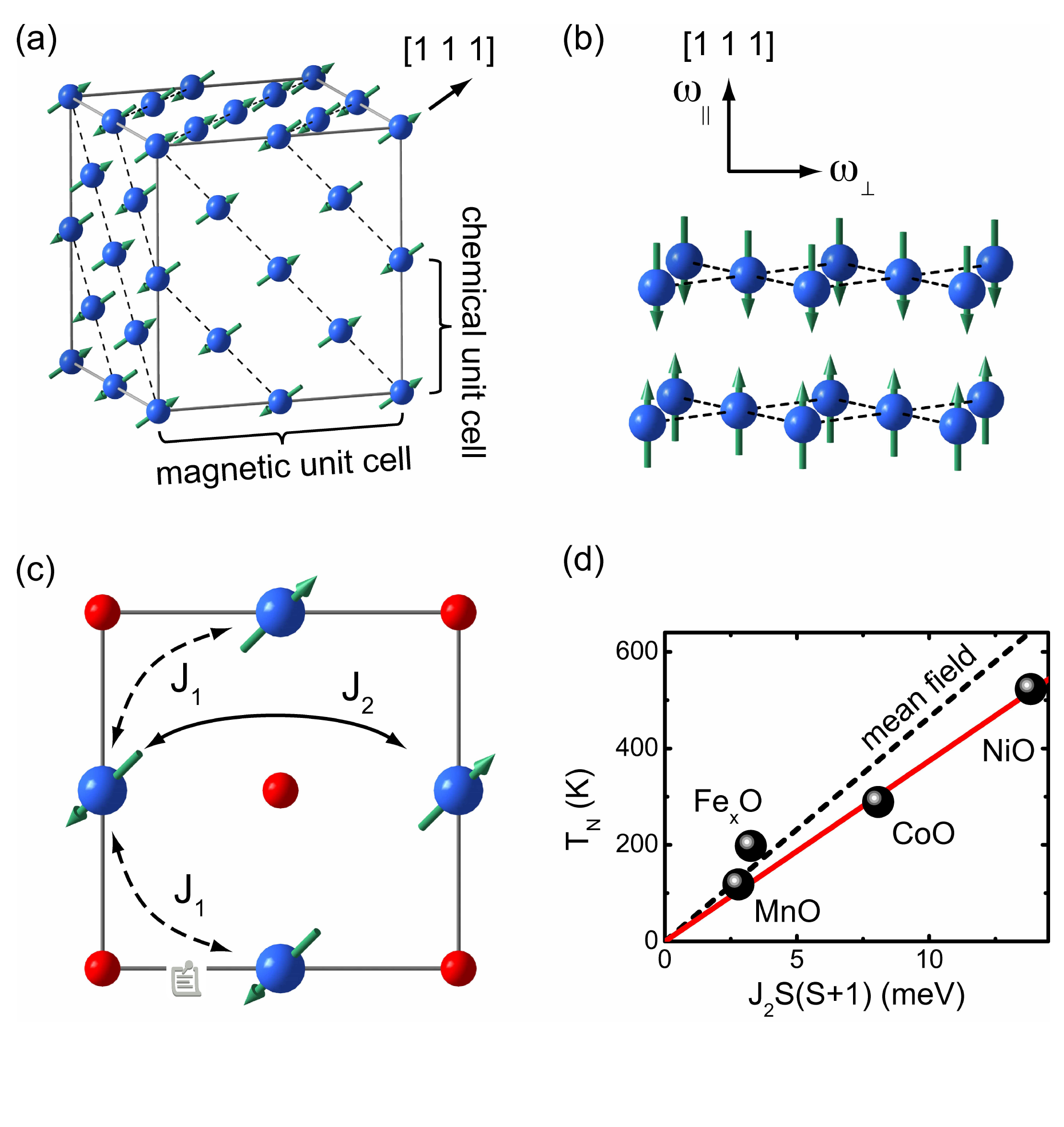}
\caption{\label{fig1}Magnetic properties of the investigated TMMOs.
(a) Magnetic unit cell showing the antiferromagnetic order along [1
1 1]. (b) Splitting into phonon modes with eigenfrequencies $\omega_{\parallel}$ and $\omega_{\perp}$. (c) Nearest-neighbour
coupling $J_1$ and dominant next-nearest neighbour coupling $J_2$.
(d) Neel temperatures vs. $J_2 S(S+1)$ using values for $J_2$ taken
from Ref.~\cite{fischer09} in comparison to the mean-field
expectation (dashed line).}
\end{figure}

TMMOs are both textbook examples for antiferromagnets governed by
superexchange in a cubic rock-salt lattice and benchmark materials
for the understanding of strongly correlated electronic systems
\cite{kunes08,zaanen85}. The magnetic structure of the TMMOs
consists of ferromagnetic planes coupled antiferromagnetically,
e.g., along [1\,1\,1] as depicted in  Fig.~\ref{fig1}(a). The
antiferromagnetic 180$^{\circ}$ next-nearest neighbor ($nnn$)
exchange $J_2$ is supposed to be the driving force of the magnetic
ordering \cite{anderson50a,roth58}, leaving the nearest-neighbor
($nn$) exchange $J_1$ frustrated, since it cannot satisfy all its
pairwise interactions [see Fig.~\ref{fig1}(c)]. In Fig.\ref{fig1}(d)
we plot the N\'eel temperatures of the TMMOs (MnO: \TN{} = 118~K
\cite{morosin70}, Fe$_{0.92}$O: $\TN{} \approx 198$~K
\cite{mccammon84,schrettle12}, CoO: $\TN{} = 289$~K
\cite{kant08,henry-la51}, and NiO: $\TN{} = 523$~K \cite{rooksby43})
as a function of $J_2 S(S+1)$, using $J_2$ values from
\cite{fischer09}, and find a linear slope of $k_B T_N/ J_2 S (S +
 1)\sim 3$ (solid line) close to the expected relation in mean-field approximation (dashed
line). In a pioneering paper Massidda et al.~\cite{massidda99}
showed that even for purely cubic TMMOs the antiferromagnetic order
is accompanied by a Born-effective-charge redistribution from
spherical to cylindrical with the antiferromagnetic axis being the
symmetry axis, e.g. [1\,1\,1]. Consequently, the cubic zone-centre
optical phonon is predicted to split into two phonon modes with
eigenfrequencies $\omega_{\parallel}$ and $\omega_{\perp}$ for light
polarized parallel and perpendicular to the cylindrical axis,
respectively [see Fig.~\ref{fig1}(b)]. The pure lattice
contributions due to deviations from the cubic symmetry in the
magnetically ordered state were estimated in the case of MnO and NiO
to be less than 1\% of the phonon eigenfrequencies
\cite{massidda99,uchiyama10}, while the exchange-driven and
experimentally observed splittings are one order of magnitude larger
\cite{chung03,rudolf08,kant10a}.

\begin{figure}[t]
\includegraphics[width=0.45\textwidth]{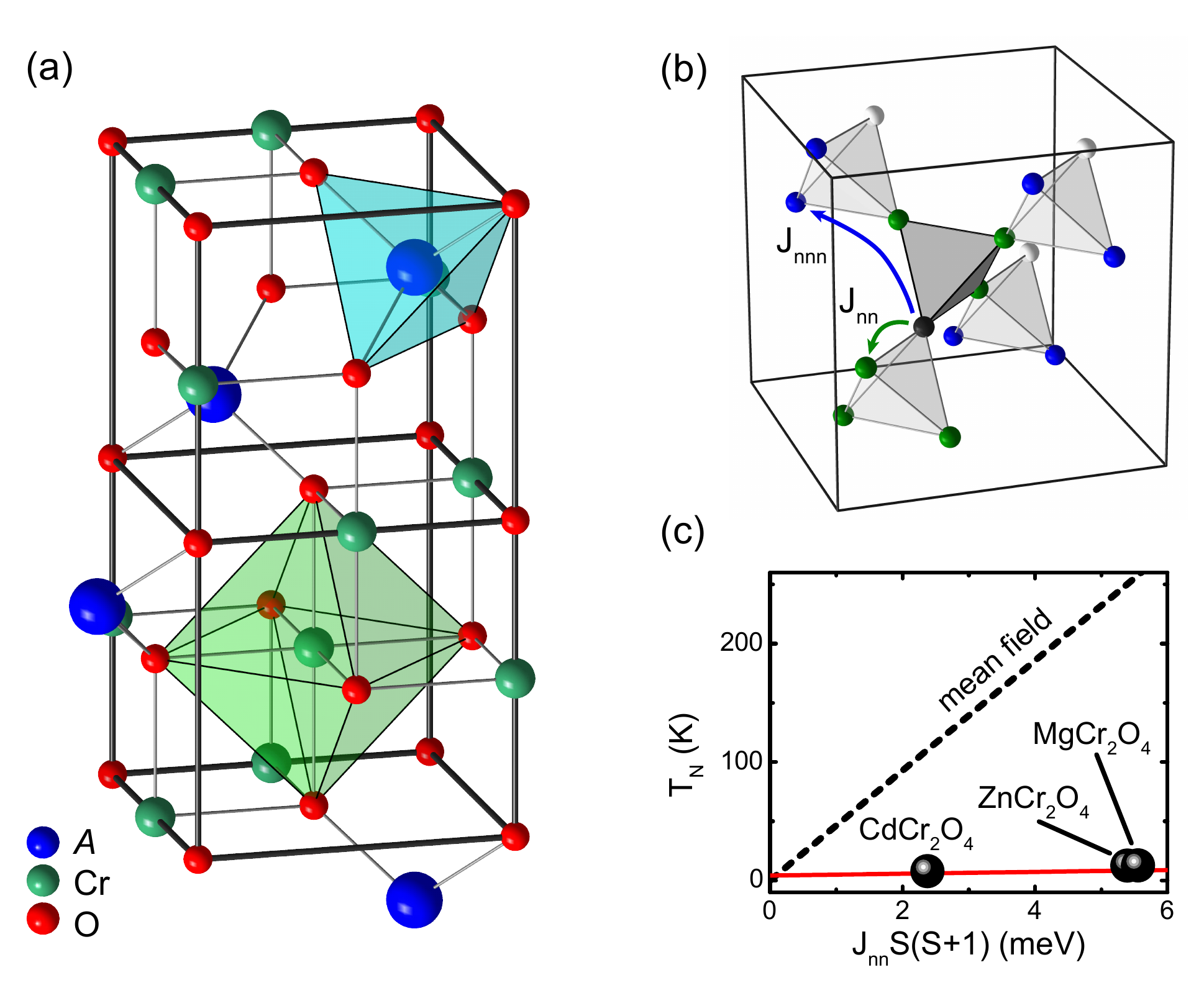}
\caption{\label{fig2} (a) Cubic spinel structure of
ACr$_{2}$O$_{4}$. Non-magnetic A-site ions are in tetrahedral and
the Cr ions in octahedral environment. (b) Nearest-neighbor and
effective further nearest-neighbor exchange paths of the Cr ions on
the pyrochlore lattice \cite{chern08,kant10a}. (c) N\'eel
temperatures vs. $J_{nn} S(S+1)$ using values for $J_{nn}$ taken
from Ref.~\cite{kant10a} in comparison to the mean-field expectation
(dashed line).}
\end{figure}

In the case of the spinel systems $A$Cr$_{2}$O$_{4}$ with
non-magnetic ions $A$ = Zn, Mg, Cd, the magnetic properties are
determined by Cr$^{3+}$ ions with spin $S = 3/2$ in octahedral
environment. The Cr sites form a pyrochlore lattice which can be
regarded as a network of corner-sharing tetrahedra [see
Fig.~\ref{fig2}(a,b)]. The inherent frustration of Heisenberg spins
on the pyrochlore lattice with antiferromagnetic $nn$ exchange
interaction $J_{nn}$ can be lifted by taking into account $nnn$
exchange coupling $J_{nnn}$ [see Fig.~\ref{fig2}(b)] \cite{chern08}
or magneto-elastic coupling leading to spin-Jahn-Teller transitions
\cite{tchernys02}, which occur at N\'eel temperatures of 12.5, 12.7
and 7.8~K  for ZnCr$_{2}$O$_{4}$, MgCr$_{2}$O$_{4}$, and
CdCr$_{2}$O$_{4}$, respectively \cite{rovers02,lee00}. These N\'eel
temperatures are low in comparison to their Curie-Weiss temperatures
of -390, -346 and -71~K \cite{rudolf07,sushkov05} and, in contrast
to the TMMOs, cannot be described in mean-field approximation using
the dominant $nn$ direct exchange constants $J_{nn}$ = 1.44, 1.48,
and 0.63~meV obtained from the analysis of the paramagnetic
susceptibilities [see Fig.~\ref{fig2}(c)] \cite{garcia-a00,kant10a}.
In the paramagnetic phase the four expected triply degenerate
optical phonons are observable in the far infrared spectra for all
three compounds \cite{lutz91}. For ZnCr$_{2}$O$_{4}$ and
CdCr$_{2}$O$_{4}$ one of these modes reportedly exhibits a
pronounced splitting into a singly and a doubly degenerate mode at
$T_N$ \cite{sushkov05,fennie06,aguilar08,kant09a}, analogously to
the TMMOs. Although the exact lattice symmetry and the spin
configuration of the magnetically ordered state are still subject of
debate \cite{chung05,lee07,ji09,kant09a}, the dominant structural
feature of the low-temperature phase is a tetragonal distortion with
an elongation along [0\,0\,1] for CdCr$_{2}$O$_{4}$ and a
contraction for ZnCr$_{2}$O$_{4}$ \cite{lee00} and MgCr$_{2}$O$_{4}$
\cite{ehrenber02}.

\begin{figure}
\includegraphics[width=0.45\textwidth]{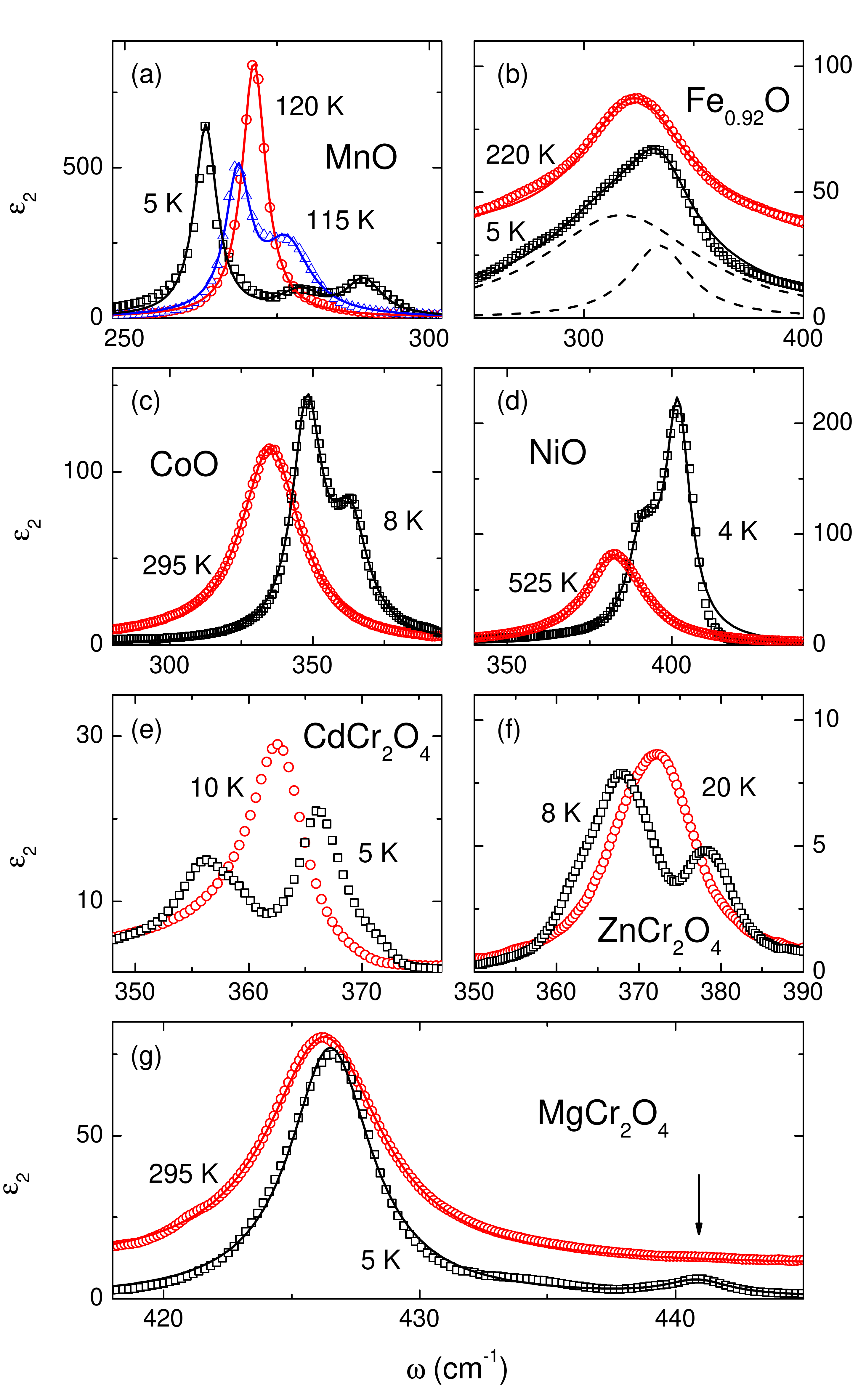}
\caption{\label{fig3}Dielectric loss of (a) MnO,  (b) Fe$_{0.92}$O,
(c) CoO, (d) NiO, (e) CdCr$_{2}$O$_{4}$, (f) ZnCr$_{2}$O$_{4}$, and
(g) MgCr$_{2}$O$_{4}$ around the TO phonon modes above and below
\TN{}, respectively. The high temperature data of Fe$_{0.92}$O and
MgCr$_2$O$_4$ are shifted upward for clarity. Lines indicate fits to
the loss peaks (see text).}
\end{figure}

The MgCr$_{2}$O$_{4}$ single crystals were grown for this study by
chemical transport in similar conditions as reported previously for
ZnCr$_{2}$O$_{4}$ \cite{kant09a}. Details for sample preparation and
characterization of the other samples have been given earlier
\cite{kant09,kant09a}. The magnetic susceptibilities of our spinel
samples \cite{kant10a} agree nicely to the ones of fully
stoichiometric samples \cite{dutton11}. In Fig.~\ref{fig3} we show
the dielectric loss functions derived via the Kramers-Kronig
relation from optical reflectivity spectra, of the transverse
optical (TO) modes for the TMMOs [(a)-(d)]
\cite{rudolf08,kant08,kant09} and the relevant phonon for the Cr
spinels [(e)-(g)] \cite{kant09a} below and above the corresponding
transition temperatures.

The splitting of the phonon mode is prominent in MnO
[Fig.~\ref{fig3}(a)] where above the N\'eel temperature (\TN{} = 118
K \cite{morosin70}) one single Lorentzian-like normal mode can be
detected. On cooling it splits right at the onset of long-range
magnetic order, where a clear shoulder appears at 115~K. Finally, at
5~K three distinct oscillators can be identified. The overall
splitting amounts to $\Delta\omega = 25.6\wn{}$, which is in
excellent agreement with neutron scattering results \cite{chung03}.
This splitting of approximately 10\% of the cubic phonon frequency
is one order of magnitude larger than what is expected from the
structural distortions \cite{massidda99}. A similar analysis was
performed for the other monoxides Fe$_{0.92}$O, CoO, and NiO. The
corresponding dielectric loss functions at low temperatures and
above \TN{} are plotted in Fig.~\ref{fig3}[(b)-(d)], respectively.
In the paramagnetic state all spectra can be well described by a
single symmetric Lorentzian line as indicated in the figure. At the
onset of magnetic order a clear anisotropy becomes apparent in all
compounds and at least two Lorentz oscillators are needed to
describe the low temperature phonon behavior. In the iron monoxide
the split phonon modes appear as an asymmetric loss peak below \TN{}
as depicted by the two dashed Lorentzian lines in
Fig.~\ref{fig3}(b), which were superposed to describe the spectrum.
The origin of the comparatively broad peaks in Fe$_{0.92}$O may be
due to strong anharmonicities and disorder in the iron deficient
structure. At the lowest measured temperatures the overall splitting
$\Delta\omega$ can be evaluated from the peak maxima of the loss
functions and amounts to $17.9\wn{}$, $14.9\wn{}$, and $-10.3\wn{}$
in Fe$_{0.92}$O, CoO, and NiO, respectively. The negative sign of
$\Delta\omega$ for NiO indicates that $\omega_{\parallel}<
\omega_{\perp}$ in agreement with a recent inelastic x-ray study
reporting a splitting of $-7.2\wn{}$ at room temperature
\cite{uchiyama10}.

Figures~\ref{fig3} [(e)-(g)] present the respective results obtained
for the chromium spinels. The case of CdCr$_{2}$O$_{4}$ resembles
the one of NiO, because the observed splitting $\Delta\omega$ =
$-10\wn{}$ of the cubic phonon mode [Fig.~\ref{fig3}(e)] leads to
$\omega_{\parallel}< \omega_{\perp}$ \cite{aguilar08,kant09a}, while
for ZnCr$_{2}$O$_{4}$ [Fig.~\ref{fig3}(f)] and MgCr$_{2}$O$_{4}$
[Fig.~\ref{fig3}(g)] we encounter the opposite situation, with
$\Delta\omega$ = 11~\wn{} \cite{sushkov05,kant09a} and
$\Delta\omega$ = 14.3~\wn{}, respectively. The size of the splitting
has previously been associated with the spin-phonon coupling effects
due to the dominant direct $nn$ exchange coupling $J_{nn}$ of the
Cr$^{3+}$ ions residing on the frustrated pyrochlore lattice. The
sign reversal of $\Delta\omega$ in the case of CdCr$_{2}$O$_{4}$
with respect to (Zn,Mg)Cr$_{2}$O$_{4}$, however, could not be
explained by this approach. It has been pointed out that this sign
reversal is in contradiction to what is expected considering that in
CdCr$_{2}$O$_{4}$ the lattice undergoes an elongation along the
tetragonal $c$-axis (leading to $\Delta\omega>0$)  while in
ZnCr$_{2}$O$_{4}$ and MgCr$_{2}$O$_{4}$ it becomes contracted
(leading to $\Delta\omega<0$) \cite{aguilar08}. Hence, we face two
scenarios, a purely structurally driven splitting and a splitting
due to spin-phonon coupling via the direct exchange coupling
$J_{nn}$, but none of the two can explain the experimental
observations for the phonon splitting in the Cr spinels.

\begin{figure}[t]
\includegraphics[width=0.5\textwidth]{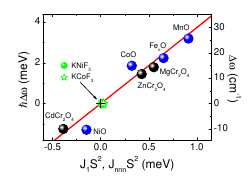}
\caption{\label{fig4} Phonon splitting $\hbar\Delta\omega$ vs.
non-dominant exchange contributions $J_1S^2$ and $J_{nnn}S^2$ for
the investigated TMMOs and Cr spinels. The solid line is a linear
fit to the experimental data of the TMMOs. The values for $J_1$ and $J_{nnn}$ are
taken from Refs.~\cite{fischer09} and \cite{kant10a}, respectively.
Values for the perovskite fluorides were taken from Ref.~\cite{axe67,balkanski67}.}
\end{figure}

Recently, Luo and coworkers \cite{luo07} proposed for the TMMOs that
the actual size of the exchange-driven phonon splitting
$\Delta\omega = \omega_{\parallel}- \omega_{\perp}$ is solely
determined in sign and magnitude by the non-dominant $nn$ exchange
$J_1$ via $\hbar\Delta\omega = \beta J_1 S^2$, while the
contributions of the dominant 180$^\circ$ $nnn$ superexchange
coupling $J_2$ are canceled. Here $S$ denotes the spin of the
transition-metal ion and $\beta$ a dimensionless factor taking into
account the dependence on lattice parameters and the vibrational
displacements \cite{luo07}. To test this prediction, we used
quantitative theoretical estimates of $J_1$ which have been obtained
only recently \cite{fischer09}. Using these values we plot in
Fig.~\ref{fig4} the experimentally observed splitting $\Delta\omega$
against the expected exchange-induced splitting $J_1S^2$ for all
investigated TMMOs. The linear dependence of both quantities
evidences that not only the size of the splitting can be
successfully described in the framework of the purely
exchange-driven scenario, but even the sign change from an
antiferromagnetic $nn$ exchange in CoO to a ferromagnetic $nn$
exchange in NiO is reflected by the inversion of the split phonon
doublet and singlet modes. From a linear fit we obtain the
dimensionless slope $\beta = 3.7$.

We extended this approach to the case of the Cr spinels. The
non-dominant $nnn$ exchange constants $J_{nnn}$ = 0.19, 0.25, and
$-0.17$~meV were obtained by the aforementioned analysis of the
paramagnetic susceptibilities for ZnCr$_{2}$O$_{4}$,
MgCr$_{2}$O$_{4}$, and CdCr$_{2}$O$_{4}$, respectively
\cite{kant10a}. Plotting the observed phonon splittings as a
function of $J_{nnn}S^2$, we found a perfect correlation with the
respective data on TMMOs in Fig.~\ref{fig4}. Note that
CdCr$_{2}$O$_{4}$ differs from the other two Cr compounds by the
reversed sign of $J_{nnn}$ (indicating an effective non-dominant
ferromagnetic coupling similar to NiO), although the absolute values
of $J_{nnn}$ are comparable for all three spinel compounds. The
agreement between the sign change of the non-dominant exchange
constant and the inversion of the split modes may resolve the
dilemma to match the observed lattice distortion and the phonon
splitting in CdCr$_{2}$O$_{4}$ mentioned above. Indeed, this finding
suggests to look at spin-phonon splitting in highly-frustrated
magnets from a new perspective and challenges the prevailing
approach to attribute the phonon-splitting to the effects of the
dominating exchange interaction \cite{sushkov05,fennie06}.

Moreover, we find not only a linear relation for the spinels, too,
but even a perfect match with the line with slope $\beta = 3.7$
obtained for the TMMOs. This result indicates that the phonon
splitting in both classes of materials originates from the same
underlying mechanism, namely the exchange-driven splitting
determined \textit{in size and sign} by the non-dominant exchange
coupling, yielding a universal proportionality factor of $~3.7$. A
universal law will have the power to predict the non-dominant
exchange splitting, if the phonon splitting is known, and vice
versa. A remarkable feature in this respect is that no
exchange-induced phonon splitting is expected, if the non-dominant
coupling is negligible, because the universal line passes through
the origin.

To comply with the situation in the TMMOs and the Cr spinels,
further systems ought to be cubic in the paramagnetic phase and only
undergo a distortion when AFM ordering sets in. Moreover, this
scenario should be valid for materials of another different
structural class. We identified corresponding materials in the class
of transition-metal perovskites, namely KCoF$_3$ and KNiF$_3$ with
corresponding Neel temperatures of 114 and 275~K \cite{okazaki61},
respectively. Most AFM perovskites are prone to show deviations from
cubic symmetry already above the N\'{e}el temperature, but in
KNiF$_3$ no deviations from cubic symmetry could be resolved even
below $T_N$ while cubic KCoF$_3$ undergoes a small tetragonal
distortion in the AFM state \cite{okazaki61}. Moreover, the above
systems are considered paradigmatic isotropic Heisenberg
antiferromagnets with negligible nnn exchange coupling
\cite{deJong74, yamaguch99, lines67, buyers71}. In addition, optical
IR studies of these two compounds did not resolve any splitting of
the three cubic triply degenerate IR active modes below $T_N$
\cite{axe67,balkanski67}. Adding the data for this class of cubic
AFM perovskites to Fig.~\ref{fig4} is clearly consistent with the
universal line derived from the rock-salt and spinel type
antiferromagnets and further supports our finding. Further materials
such as, for example, CrN \cite{filippetti00} or MnSe
\cite{popovic06} should be reexamined with respect to an
exchange-induced phonon splitting.

Although the linear slope has been predicted in a straightforward
manner for the TMMOs in \cite{luo07}, a detailed analysis of direct
and indirect exchange interactions and their dependence on the
respective phonon-modulated exchange paths of all three classes of
antiferromagnets appears to be necessary \cite{harrison07}. In
particular, the role of non-dominant couplings has not been treated
in previous theoretical approaches \cite{massidda99,fennie06}, and
Uchiyama suggested that charge-transfer processes and Jahn-Teller
effects play an important role in the TMMOs \cite{uchiyama12}. In
this respect we would like to point out three routes, which could
provide further insight into the observed phenomenological relation:
(i) calculations similar to the ones performed for the TMMOs
\cite{massidda99} and the spinels \cite{fennie06}, where the
exchange-driven phonon-splitting was assumed to stem from the
dominant exchange coupling, should be undertaken for the perovskite
systems, too. (ii) Experimentally, it might be possible to move
along the universal curve by changing the size of the non-dominant
exchange coupling by an external parameter (such as pressure or
magnetic field), to determine this change and track down the
respective phonon splitting in the optical experiment. The pressure
dependence of the exchange coupling constants has been predicted for
the TMMOs \cite{fischer09}. The non-dominant coupling will increase
under pressure and, e.g. in MnO one may expect a significantly
enhanced phonon-splitting under hydrostatic pressure for the
universal line. (iii) Finally, we want to mention that the influence
of pressure on the N\'{e}el temperature and magnetic excitations has
previously been studied for some TMMOs and led to the empirical law
$-\partial \ln J/\partial \ln V \simeq$ 10/3 for the volume
dependence of superexchange interactions \cite{bloch66,johnson74}.

In summary, we found a universal linear relation of the observed
splitting of the zone-centre optical phonon and the non-dominant
exchange couplings in transition-metal monoxides and frustrated
Cr-oxide spinels, yielding a dimensionless slope of 3.7. For the
TMMOs our results are in agreement with the predictions for an
exchange-driven phonon splitting by the non-dominant $nn$ exchange
coupling. The universal linear relation not only correctly describes
the size of the splitting, but even the sign change in the
non-dominant exchange coupling is compatible with the inversion of
the phonon modes with eigenfrequencies $\omega_{\parallel}$ and
$\omega_{\perp}$. In systems with negligible non-dominant coupling
such as the perovskites KCoF$_3$ and KNiF$_3$ the universal law
predicts no splitting in agreement with the experimental
observation. This paves a new way for understanding spin-phonon
coupling effects in antiferromagnets and, in particular, their role
in releasing frustration in highly frustrated magnets.

It is a pleasure to thank Walt Harrison and H. Uchiyama for
stimulating discussions and a critical reading of the manuscript,
and we acknowledge fruitful discusssions with F Rivadulla, C.
Batista, G.-W. Chern, and O. Tschernyshyov. This work was supported
by the Deutsche Forschungsgemeinschaft via TRR 80 (Augsburg-Munich).

%\bibliography{database}


\end{document}